\documentclass{elsart3} 
\usepackage{graphicx}

\usepackage{amssymb}

\begin{document}

\begin{frontmatter}



\title{Collective dynamics in phospholipid bilayers investigated
by inelastic neutron scattering: Exploring the dynamics of
biological membranes with neutrons}


\author[ILL,FZJ]{M.C.~Rheinst\"{a}dter}
\author[IRP]{C.~Ollinger}
\author[ILL]{G.~Fragneto}
\author[IRP]{T.~Salditt}

\address[ILL]{Institut Laue-Langevin, BP 156, 6 rue Jules Horowitz, 38042
Grenoble Cedex 9, France}

\address[FZJ]{Institut f\"ur Festk\"orperforschung, FZ-J\"{u}lich, 52425
J\"{u}lich, Germany}

\address[IRP]{Institute for X-ray Physics,
Georg-August-Universit\"at G\"ottingen, Geiststra\ss e 11, 37073
G\"ottingen, Germany}

\begin{abstract}
We present the first inelastic neutron scattering study of the
short wavelength dynamics in a phospholipid bilayer. We show that
inelastic neutron scattering using a triple-axis spectrometer at
the high flux reactor of the ILL yields the necessary resolution
and signal to determine the dynamics of model membranes. The
results can quantitatively be compared to recent Molecular
Dynamics simulations. Reflectivity, in-plane correlations and the
corresponding dynamics can be measured simultaneously to gain a
maximum amount of information. With this method, dispersion
relations can be measured with a high energy resolution. Structure
and dynamics in phospholipid bilayers, and the relation between
them, can be studied on a molecular length scale.
\end{abstract}

\end{frontmatter}

Motions in biological membranes span a wide range of length and
time scales \cite{Lipowsky1995}. Short wavelength density
fluctuations in the plane of the bilayer (as opposed to bilayer
bending modes) are likely to play a key role in the transport of
small molecules across the bilayer \cite{Paula1996}, but have
received little experimental attention. While the structure of
lipid bilayers has been the object of several studies in the last
decades, dynamical properties even of simple model systems such as
1,2-dimyristoyl-sn-glycero-3-phoshatidylcholine (DMPC) remain
largely unknown. A more quantitative understanding of short range
molecular motions in a lipid bilayer is of fundamental interest in
membrane biophysics. Experimental results can be linked to modern
Molecular Dynamics (MD) simulations \cite{Tarek2002}. In this
context, Chen {\em et al.} have recently presented a seminal
measurement of the dispersion relation of the acyl chain
fluctuations in lipids \cite{Chen2001} by inelastic X-ray
scattering (IXS). As an important result, Chen {\em et al.} found
a minimum in the dispersion relation at about Q$_0$=1.4
\AA$^{-1}$, at the position of the static structure factor maximum
S(Q$_{xy}$). This soft mode was hypothetically linked to transport
phenomena across the bilayer.
Inelastic neutron scattering using a triple-axis spectrometer
should offer a much better resolution and avoid the danger of
radiation damages in the sample, but has to date not been applied
to study the dynamics of lipid bilayers. A further important
advantage of INS over IXS is the fact that different parts of the
bilayer can be probed by use of selective deuteration. We carried
out the first inelastic neutron scattering experiments in model
membranes employing the cold triple-axis spectrometer IN12 at the
high flux reactor at the ILL. In this study we used fully
deuterated chains DMPC -d54 (Avanti lipids, Alabama) in D$_2$O to
minimize incoherent background. We report on experimental details
and show examples of neutron diffraction and inelastic neutron
scattering in deuterated DMPC bilayers.
\begin{figure}
\centering
\resizebox{0.85\columnwidth}{!}{\rotatebox{0}{\includegraphics{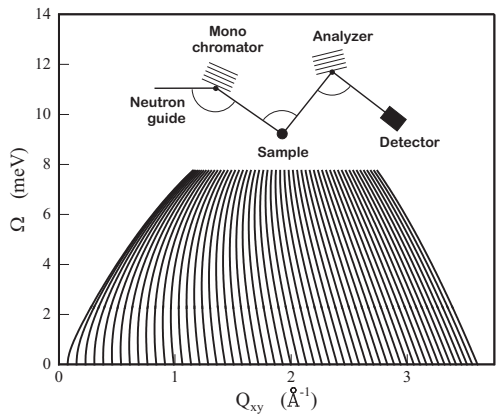}}}
\caption[]{Accessible (Q,$\Omega$) range for a fixed energy of the
scattered beam of E$_f$=10 meV. In the insert, a schematic of a
triple-axis spectrometer is shown.} \label{qrange.eps}
\end{figure}

From a solution of  DMPC -d54 (deuterated
1,2-dimyristoyl-sn-glycero-3-phoshatidylcholine) in TFE/chloroform
(1:1) with a concentration of 40 mg/ml, we prepared highly
oriented membrane stacks on 2'' silicon wafers (thickness 300
$\mu$m)\cite{Muenster1999}. Ten of these wafers were stacked on
top of and aligned with respect to each other to create a
'sandwich sample' consisting of several thousands of highly
oriented lipid bilayers with a total mass of about 300 mg of
deuterated DMPC. Aluminum spacers between the substrates allow to
let the heavy water vapor in between the wafers to hydrate the
bilayers. In rocking scans we find a mosaicity of the sandwich of
$\Delta\omega$=0.6 deg.

\begin{figure} \centering
\resizebox{0.90\columnwidth}{!}{\rotatebox{0}{\includegraphics{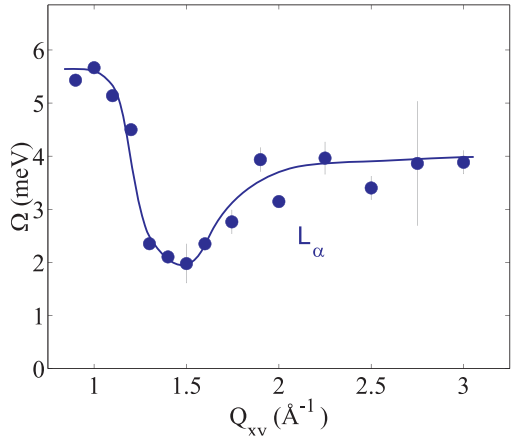}}}
\caption[]{Dispersion relation in the fluid phase of the lipid
bilayer. Note the minimum at the maximum of the static structure
factor S(Q$_{xy}$) shown in Fig. \ref{kettenpeak}.}
\label{dispersion}
\end{figure}

\begin{figure} \centering
\resizebox{0.90\columnwidth}{!}{\rotatebox{0}{\includegraphics{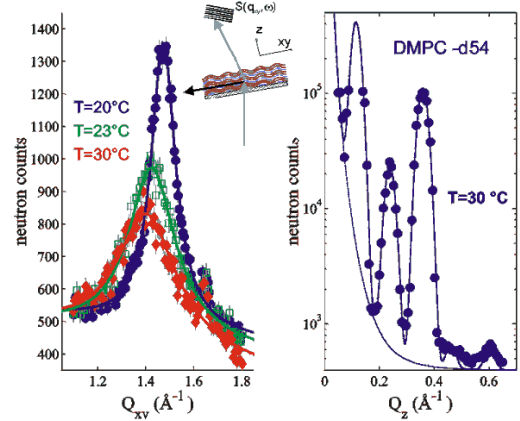}}}
\caption[]{The inter-acyl-chain correlation peak (left) and
reflectivity (right). The insert schematically shows the
orientation of the sample with respect to the incoming beam for
the in-plane measurements. Solid lines are fits after the
Lorentzian- (chain-peak) and Gaussian-model (reflectivity).}
\label{kettenpeak}
\end{figure}

The insert in Fig. \ref{qrange.eps} shows a schematic of a
triple-axis spectrometer. By varying the three axes of the
instrument, the axes of rotation of the monochromator, the sample
and the analyzer, it is possible to change the wavevectors k$_i$
and k$_f$ and the energies E$_i$ and E$_f$ of the incident and the
scattered beam, respectively. The accessible ({\bf Q},$\Omega$)
range for a fixed energy of the scattered beam E$_f$ of 10 meV is
shown in Fig. \ref{qrange.eps}. It is limited by the range of
incident neutron energies offered by the neutron guide as well as
by mechanical restrictions of the spectrometer. The instrumental
energy resolution in this configuration is $\Delta\Omega$=500
$\mu$eV. By choosing smaller incident energies and energy
transfers the energy resolution can be enhanced. By rotating the
sample the scattering vector {\bf Q} can be placed either within
the plane of the membranes (Q$_{xy}$) to measure inter-chain
correlations or perpendicular to it (Q$_z$) to measure
reflectivity (see insert in Fig. \ref{kettenpeak}). The use of a
triple-axis spectrometer offers the unique possibility to measure
reflectivity, the static structure factor S(Q$_{xy}$) in the plane
of the membranes and in plane dynamics (S(Q$_{xy},\Omega$)) on the
same instrument in the same run. To control temperature and the
degree of hydration during the experiment the sample was kept in a
humidity chamber. Inside the chamber there was a heavy water
reservoir, thermally isolated against the sample. By adjusting the
temperature of the chamber and independently the temperature of
the reservoir using two Haake bath controllers, the humidity
inside can be controlled. Close to the sample there were a
humidity sensor and a platinum resistor installed. IN12 is
equipped with a vertically focusing graphite monochromator and a
vertically and horizontally curved graphite analyzer. The flux at
the sample position is of the order of 5$\cdot$10$^6$
n/(cm$^2\cdot$ s). For the inelastic measurements the counting
times were about five minutes per point.
\begin{figure}
\centering
\resizebox{0.90\columnwidth}{!}{\rotatebox{0}{\includegraphics{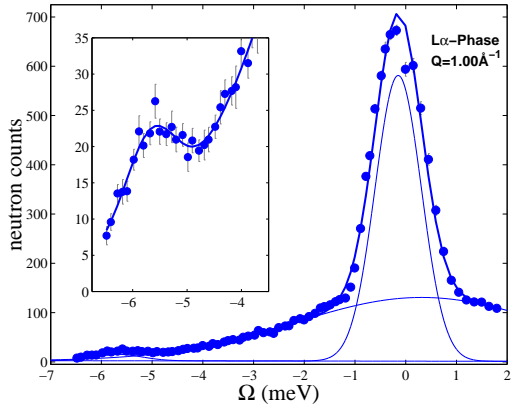}}}
\caption[]{Energy scan at Q=1.0 \AA$^{-1}$ and T=30 $^{\circ}$C.
The insert shows the excitation of the DMPC-layers in
magnification. Solid lines are guides to the eye.}
\label{q100_final.eps}
\end{figure}
Figure \ref{q100_final.eps} shows an energy-scan at Q$_{xy}$=1.0
\AA$^{-1}$ and T=30 $^{\circ}$C, in the fluid phase of the DMPC
bilayer.  The signal consists of a sharp central peak and a broad
contribution, centered at $\Omega$=0 meV. The solid lines are
guides to the eye. At $\Omega$=-5.5 meV, the inelastic signal of
the membrane is observed in form of a pronounced peak with a width
(FWHM) of about $\Delta\Omega$=1 meV. Figure \ref{dispersion}
shows the corresponding dispersion relation in the fluid L$\alpha$
phase at T=30 $^{\circ}$C as measured by several constant-Q scans.
The measurements can quantitatively be compared  to corresponding
Molecular Dynamics simulations by Tarek {\em et
al.\@}\cite{Tarek2002}.
Figure \ref{kettenpeak} shows the corresponding inter-acyl-chain
peak for temperatures below and above the gel-liquid phase
transition (phase transition temperature T$_c$=21.5 $^{\circ}$C)
and the reflectivity at T=30 $^{\circ}$C. When heating from the
more rigid gel-phase into the fluid phase, the inter-chain peak
broadens drastically (decreasing correlation length $\xi_{xy}$)
and the peak center shifts to smaller Q$_{xy}$-values (larger
average next neighbor distance and smaller packing density). The
reflectivity shows five well-developed Bragg peaks giving
information about the distance of the layers in $z$-direction.
From the structure factors of the Bragg peaks, the $z$-profile of
the bilayer can be modelled. At T=30 $^{\circ}$C we find a
next-neighbor distance of the acyl chains of d$_{xy}$=4.5 \AA\ and
a correlation length of $\xi_{xy}$=11 \AA. From the reflectivity
there is a distance between the bilayers of d$_z$=55 \AA. The
combination of diffraction and inelastic measurements leads to a
complete picture of structure and dynamics of model membranes on a
molecular length scale. This will be especially useful in
temperature dependent measurements in the range of the gel-liquid
phase transition to study the development of structure and
dynamics and the relation between them. This work has been funded
by the German Research Ministry under contract number 05-300-CJB-6
(MCR).




\end{document}